\documentclass[aps,twocolumn,showpacs,superscriptaddress]{revtex4-1}
\usepackage{latexsym,amssymb}
\usepackage{graphicx}
\usepackage{amsmath}
\usepackage{epstopdf}
\usepackage{placeins}
\usepackage{textcomp}
\usepackage{ulem}

\begin{document}
\title{One-dimensional matter waves as a multi-state bit}

\author{J. Giacomelli}
\affiliation{Sace, Piazza Poli 37/42, 00187 Roma, Italy}

\begin{abstract}
We design a technique to control the position of a localized matter wave. Our system is composed by a two counter-phased periodic potentials and a third optical lattice which can be chosen to be either periodic or disordered. The only control needed on the system is a three-state switch that allows the instantaneous selection of the desired potential. We show that this framework is robust and the multi-state bit behavior can be observed under generic hypothesis.
\end{abstract}


\maketitle

\section{Introduction}
\label{intro}
Nowadays Bose Einstein condensates (BEC) \cite{Cornell1995, Ketterle1995, Cornell1996, Inguscio2001} are routinely used in combination with optical potentials in order to have a direct access to the fundamental quantum behaviors on a macroscopic scale. The state of art in this field offers a wide range of possibilities in terms of manipulation over these systems \cite{Modugno2001, Ferlaino2002, Henderson2009, Abdullaev2008, Girardeau2001, Zhang2008, Cataliotti2003, Fort2003, Fallani2004, Ferlaino2004, meyrath2005, fallani2008, shapiro2012, sp2010, gmodugno2010, damski2003, lye2005, fort2005, clement2006} and there is a deep knowledge of the expected behaviors from the theoretical side.

The dimensionality of the system can be reduced by flattening the BEC (effective $2D$ system \cite{Ferlaino2004}) or elongating it (effective $1D$ system \cite{Cataliotti2003, Fort2003, Fallani2004}). In the $1D$ case, interesting boundary conditions can be realized: the elongated BEC can be trapped in a box \cite{meyrath2005}, in a torus \cite{Ramanathan2011} or in an harmonic trap \cite{Girardeau2001, Zhang2008}, among other possibilities \cite{Abdullaev2008, Henderson2009}.

Many different optical potentials can be realized for this system. Without any presumption of being exhaustive, we recall the possibility of generating both periodic \cite{Cataliotti2003, Fort2003, Fallani2004, Ferlaino2004} and disordered \cite{fallani2008, shapiro2012, sp2010, gmodugno2010, damski2003, lye2005, fort2005, clement2006, modugno2006, falco2010, giacomelli2014} lattices. The latter family of potentials has been employed in order to observe Anderson localization phenomena \cite{fallani2008, shapiro2012, lye2005, fort2005, giacomelli2014}. $1D$ speckle potential in particular have been the object of an intensive study in recent years, both from theoretical and experimental side. The localization properties of a speckle system have been investigated both in infinitely extended \cite{shapiro2012, sp2010} and box bounded systems \cite{falco2010, giacomelli2014}, showing that the finite length case can have an even stronger degree of localization compared with the infinite length case, under the proper conditions \cite{giacomelli2014}. In addition to the wide selection of feasible optical potentials, we recall the recent possibility of painting an arbitrary shape time$-$averaged optical dipole potential \cite{Henderson2009}.

Finally, the Fano-Feshbach resonances \cite{Khaykovich2002, Strecker2002, Roati2007} can be employed to lessen or even eliminate the non linear effects of the self$-$interaction, leading the dynamics of the system to be ruled by a linear Schr\"odinger equation.

This strong degree of control over a quantum system allows for the research of technological applications. In particular, investigations in using matter wave as quantum switches or quantum information device has been done in recent years \cite{Ahufinger2007, Modugno2014}. In this article we propose a general technique to employ a $1D$ BEC, either self-interacting or not, as a multi-state bit, by a proper temporal alternation of three optical potentials. This design is completely new to the best of our knowledge and it is the first example of BEC used as a classical multi-state bit. The proposed technique is robust and can be applied under a range of different specifications, both in the box and the torus cases. Also the number of states is an arbitrary choice.

The article is organized as follows. In Sec.~\ref{intrononint} we define the general features of our system.
In Sec.~\ref{examplesec} we discuss the way to employ this system as a multi-state bit. The robustness of the system is discussed in Sec.~\ref{robustsec}, where different possible implementations are compared. Our results are summarized in~Sec.~\ref{secend}.

\section{Model and methods}
\label{intrononint}
Let us consider a non interacting matter wave in a 1D optical potential which can be selected amongst three possible choices.  The system is finite and its length is $L$. The Hamiltonian of the system can be written in a dimensionless form as
\begin{equation}
\hat{H} = -\frac{d^2}{d x^2} + \sum_{k=1}^3\mathbf{1}_{\{k=c\}}v_k(x,s_k),
\label{hamiltonian}
\end{equation}
where $c\in\{1,2,3\}$ is the external choice of the potential and $v_i$ are three potentials described below. We consider also a zeroth case ($c=0$) which is not selectable during the time evolution of the system but it is used just to set the initial conditions.  The dimensionless hamiltonian in Eq. (\ref{hamiltonian}) is rescaled by an energy value $E_\xi=\frac{\hbar^2}{2m\xi^2}$ with $\xi\simeq 1\mu m$. $E_{\xi}$ is related to $v_1(x,s_1)$ and defined together with it.
We want to study the dynamics of this system, under the hypotesis that $c$ can be changed instantaneously. This is reasonable considering that $v_k(x,s_k)$ can be optically generated. When considering the time evolution and the presence of self-interaction, the system is fully described by the Schr\"odinger equation
\begin{equation}
i\frac{\partial}{\partial \tau}\psi = \left[\hat{H}+2\alpha\beta\frac{\vert\psi\vert^2}{\sigma^2}+ \alpha\left(\sigma^2+\sigma^{-2}\right)\right]\psi,
\label{schr}
\end{equation}
where $\tau = E_\xi t/\hbar$ is a rescaled dimensionless time and $\sigma^2(x,\tau)=\sqrt{1+\beta\vert\psi(x,\tau)\vert^2}$. $\alpha$ and $\beta$ are defined and fully specified in Appendix \ref{appnpse}. The nonlinear terms describe the self interaction of the Bose Einstein Condensate (BEC) that can be used in order to realize the system. Eq. (\ref{schr}) is an effective 1D model known as non-polynomial Schr\"odinger equation (NPSE) \cite{Salasnich2002}. This is obtained from the Gross-Pitaevskii 3D equation \cite{Stringari2003} in order to provide an approximate description for the BEC dynamics under radial confinement. Nowadays, the self interaction can be chosen to be repulsive ($\beta>0$) \cite{wu2001, smerzi2002}, attractive ($\beta<0$) \cite{Khaykovich2002, Strecker2002} or absent ($\beta=0$) \cite{Roati2007}, depending on the experimental settings. The considered potentials are defined in the following.
\begin{itemize}
\item[$(v_1)$] In Section \ref{examplesec} we explain how $v_1$ can be employed in order to keep $\vert\psi(x,\tau)\vert^2$ stable over time. To this end, we consider two possibilities: a disodered potential $v_1^{d}$ and a periodic potential $v_1^{p}$. 
 
\item[$(v_1^{d})$] We consider an optical speckle $v_{1}^d(x)=V_0 v(x/\xi)$, with  intensity $V_0 = \left\langle v_1\right\rangle$ and autocorrelation length $\xi$ \cite{goodman2005,modugno2006}. The probability distribution of $v(x)$ is $e^{-v}$. Moreover it holds that
\begin{equation}
\langle v(y)v(y+x)\rangle_y = 1+sinc^2\left(\frac{x}{\xi}\right)\nonumber
\end{equation}
Optical speckle is obtained by transmission of a laser
beam through a medium with a random phase profile, such as
a ground glass plate. The resulting complex electric field
is a sum of independent random variables and forms a Gaussian process. Atoms experience a random potential proportional to the intensity of the field. $V_0$ can be either positive or negative, the potential resulting in a series of barriers or wells. However, in both cases it is possible to observe Anderson localization phenomena \cite{falco2010, giacomelli2014}. 
The autocorrelation length $\xi$ represents a natural  scale for the system and $E_{\xi}=\hbar^2/2m\xi^2$ is the corresponding energy scale. We define
\begin{equation}
v_1^d(x,s_1) = s_1v(x)
\end{equation}
where $s_1=V_0/E_{\xi}$ is a rescaled dimensionless intensity.
The speckle pattern can be generated numerically as discussed in \cite{modugno2006} (and references therein).
\item[$(v_1^{p})$]   This potential must be smooth and periodic:
\begin{equation}
v_1^p(x,s_1) = s_1f\left(\textrm{mod}\left(x,\frac{\Delta}{2}\right)\right),
\end{equation}
where $\Delta=L/N$ ($N\in\mathbb{N}$), $f(\Delta/2+x)=f(\Delta/2-x)$ and $df(x)/dx=0\Leftrightarrow\textrm{mod}(x,\Delta/2)=0$. $\textrm{mod}(a,b)$ is the remainder of the division $a/b$.\newline\indent
In section \ref{robustsec} we consider $v_1^p(x,s_1) = s_1cos(\frac{4\pi}{\Delta} x)$ as a realistic case.
\item[$(v_2)$] $v_2$ can be obtained from $v_1^p$ by doubling the period and considering a different amplitude $s_2$, which is a parameter independent from $s_1$. 
\begin{equation}
v_2(x,s_2) = s_2f\left(\textrm{mod}(x,\Delta)\right),
\label{eqv2}
\end{equation}
where the same requirements described above holds. In Section \ref{robustsec} we consider $v_2(x,s_1) = s_2cos(\frac{2\pi}{\Delta} x)$ as a realisitic case.

\item[$(v_3)$]  Also the third potential is smooth and periodic, in antiphase with $v_2(x,s_2)$:
\begin{equation}
v_3(x,s_3) = s_3f\left(\textrm{mod}(x,\Delta)+\frac{\Delta}{2}\right)
\label{eqv3}
\end{equation}
In the following we will always consider $s_2=s_3$ only.
\item[$(v_0)$] The initial condition $\psi(x,\tau=0)$ must be localized around $x_0$ such that $\textrm{mod}(x_0,\Delta)=\Delta/4$. This can be achieved forcing the BEC to the ground state of a properly chosen optical potential $v_0(x,s_0)$. In Section \ref{examplesec} we consider
\begin{equation}
v_0(x,s_0)=s_0 \cos\left(\frac{4\pi}{\Delta} x\right)+\omega^2\left(x-\frac{L}{2}-\frac{\Delta}{4}\right)^2
\end{equation}
where $\omega^2$ is a constant dimensioned as $\textrm{length}^{-2}$ and valued as $\vert L^{-1}\vert$.
\end{itemize}
In the next section we show that the system described above acts as a multi-state bit under two alternative boundary conditions: box and torus.
\subsection{Measure of the system stability}
\label{dprpar}
In Section \ref{examplesec} we describe a method to control the localization position $x_{loc}$ of the matter wave $\psi$ by changing the $c$ value with a proper timing. Hence we are interested to prevent the spatial expansion of $\psi$, in order to be able to measure $x_{loc}(\tau)$ even for $\tau\gg 0$. The Participation Ratio (PR) is commonly used in literature as a measure of the localization degree \cite{Evers2000, giacomelli2014}.
\begin{equation}
\label{eqpr}
PR\left\lbrack\psi\right\rbrack =\frac{1}{L\int_{L} dx \left\vert\psi(x)\right\vert^4}
\end{equation} 
We introduce the following quantity in order to compare the $PR$ value measured during the evolution of the system with the initial one.
\begin{equation}
\label{eqdpr}
DPR(\tau)=\frac{PR\left\lbrack\psi(x,\tau)\right\rbrack}{PR\left\lbrack\psi(x,0)\right\rbrack}
\end{equation} 
 The measure of $x_{loc}(\tau)$ becomes more difficult and less precise at increasing $DPR(\tau)$ values. In our system $x_{loc}$ is clearly measurable when $DPR\lesssim 10$ while it cannot be defined nor observed anymore when $DPR\gtrsim 20$.     

\section{How to use the system as a multi-state bit}
\label{examplesec}
\begin{figure}[hbt]
\begin{center}
\includegraphics[width=1\columnwidth]{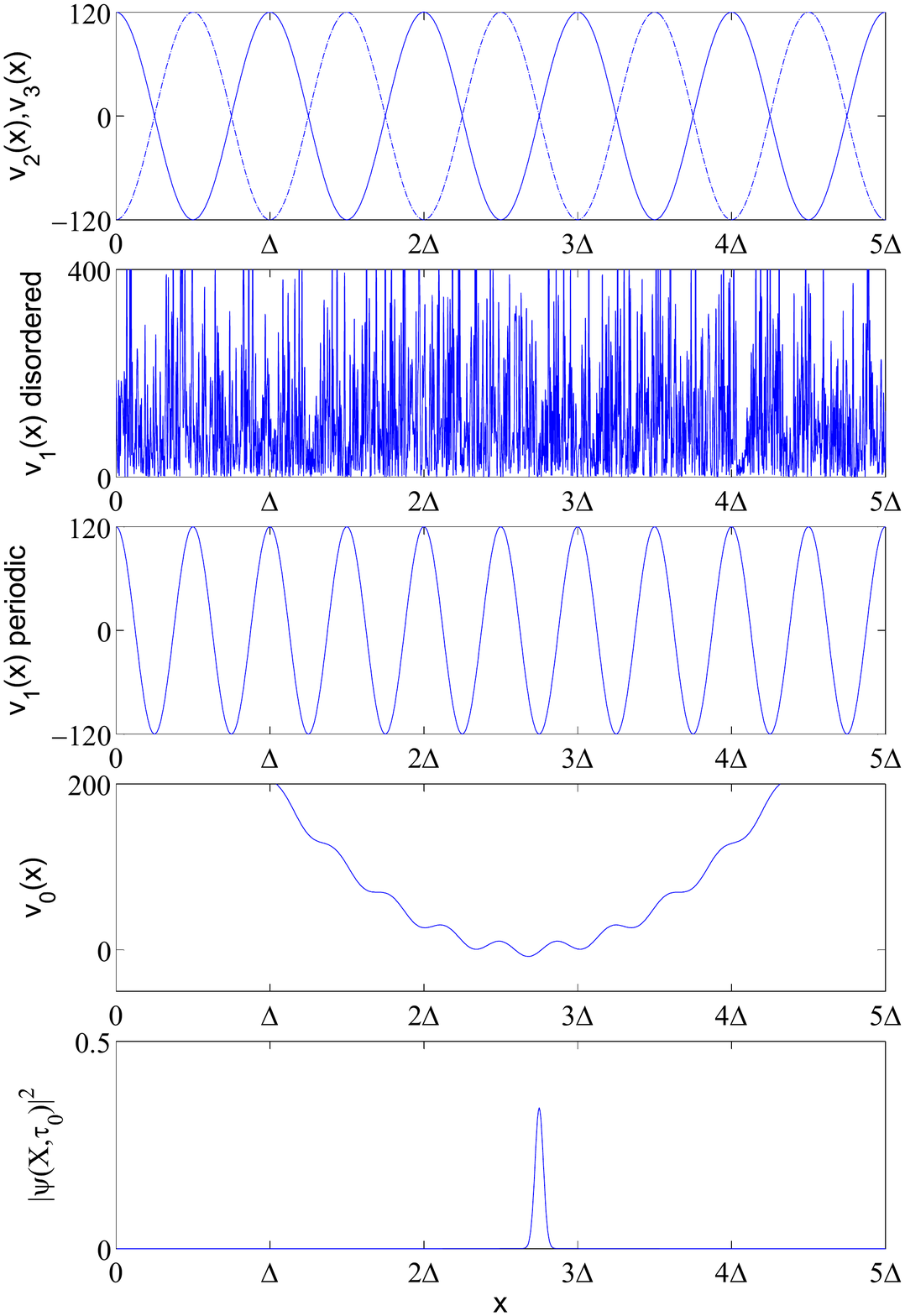} 
\end{center}
\caption{A possible specification of the system described in Section \ref{intrononint}: $N=5$ and $v_0=v_1=v_2=v_3=120$ . From the top to the bottom: the potentials $v_2$ (solid line) and $v_3$  (dotted line); the speckle potential $v_1^d$; the periodic potential $v_1^p$; the potential $v_0$; the initial density profile $\vert\psi(x,\tau_0)\vert^2$.}
\label{figexample_v}
\end{figure}
A multi-state bit can assume a state chosen from a discrete and finite set. We can conventionally define this set of states by partitioning the system into $2N=2L/\Delta$ intervals and labeling each interval with a number $n\in\{1,\dots,2N\}$. We want to perform three basic operations on our system in order to consider it a true multi-state bit: writing information, keeping memory of it over an arbitrary time lapse, reading it again. 
\subsection{Reading information from the position of a localized matter wave}
Using a matter wave allows us to measure the density profile directly. Since the chosen initial condition of the system is a localized state $\psi(x,\tau_0)$, the most of the density is concentrated in a small region. If $\textrm{DPR}(\tau)\lesssim 10$ while the system evolves, we can associate a number $n_{j}$ to every instant $\tau$ by measuring the density profile:
\begin{equation}
\begin{array}{c}\varphi_{R}\left(\psi\vert\tau\right) = \sum_{j=1}^{2N}n_ j\mathbf{1}_{\left\{j-1\leq\frac{2x_{loc}(\tau)}{\Delta}<j\right\}}\\ 
\textrm{with }x_{loc}(\tau) = \underset{x\in[0,L]}{\textrm{argmax}}\vert\psi(x,\tau)\vert^2,\end{array}
\label{eqread}
\end{equation}
Eq. (\ref{eqread}) allows us to read the information stored in our system. In the next paragraph we discuss how to write information in the system (using operators $\varphi_{+}$ and $\varphi_{-}$) and how to store the information over a time lapse $\delta$ keeping $n_{\tau}$ constant (using operator $\varphi_{\delta}$). The convenient choice of $n_{j}$ depends on the boundary conditions. In case of box boundary conditions we choose
\begin{equation}
n_j = \frac{j+1}{2}\textbf{1}_{\{\textrm{mod}(j,2)=1\}}+\left(2N-\frac{j}{2}+1\right)\textbf{1}_{\{\textrm{mod}(j,2)=0\}}
\label{eqnbox}
\end{equation} 
 In case of toroidal boundary conditions we choose
\begin{equation}
n_j = \frac{j+\textrm{mod}(j,2)}{2}
\label{eqntorus}
\end{equation} 
We discuss the reason of this choices in paragraph \ref{bound}. In our example ($N=5$) eq. (\ref{eqnbox}) leads to 
\begin{equation}
\underline{n}=(1,10,2,9,3,8,4,7,5,6)\nonumber 
\end{equation}
and eq. (\ref{eqntorus}) leads to 
\begin{equation}
\underline{n}=(1,1,2,2,3,3,4,4,5,5)\nonumber
\end{equation}
as shown in Figure \ref{figbox}. 
\subsection{Writing and maintaining information in the system} 
As discussed in Section \ref{intrononint}, the only way that we have to influence the system is switching $c_{\tau}$ from a value to another. We want to use this possibility to define three actions which affect the system as follows:
\begin{eqnarray}
\varphi_{R}\left[\varphi_{+}(\psi\vert\tau)\vert \tau+\epsilon\right] &=& \varphi_{R}\left(\psi\vert\tau\right)+1\\
\varphi_{R}\left[\varphi_{-}(\psi\vert\tau)\vert \tau+\epsilon\right] &=& \varphi_{R}\left(\psi\vert\tau\right)-1\\
\varphi_{R}\left[\varphi_{\delta}(\psi\vert\tau)\vert \tau+\delta\right] &=& \varphi_{R}\left(\psi\vert\tau\right)
\end{eqnarray}
where $\epsilon$ is the time interval necessary to apply the operators $\varphi_{\pm}$ and $\delta$ is a time interval over which the information has to be stored in the system. As explained in paragraph \ref{bound}, the boundary conditions affect the definition of the $\varphi_{+}$ and $\varphi_{-}$ operators.
In case of box conditions we have
\begin{equation}
\label{pmbox}
\varphi_{R}\left[\varphi_{\pm}(\psi\vert\tau)\vert \tau+\epsilon\right] = \textrm{mod}[\varphi_{R}\left(\psi\vert\tau\right)-1\pm 1,2N]+1
\end{equation}
In case of toroidal conditions we have
\begin{equation}
\label{pmtorus}
\varphi_{R}\left[\varphi_{\pm}(\psi\vert\tau)\vert \tau+\epsilon\right] = \textrm{mod}[\varphi_{R}\left(\psi\vert\tau\right)-1\pm 1,N]+1
\end{equation}

\subsubsection{Definition of $\varphi_{\delta}$}
\label{mempar}
The definition of $\varphi_{\delta}(\cdot\vert\tau)$ is based on different principles in case we use $v_1^d$ or $v_1^p$. \newline\indent
In case we use a disordered potential $v_1^d$, it can cause the Anderson localization of the system and it inhibits any transport phenomena. Hence, provided that the disordered potential amplitude is big enough, any localized matter wave $\psi(x,\tau_0)$ should remain localized at the same position when observed in $\tau_0+\delta$. \newline\indent
In case we use a periodic potential $v_1^p$, it can inhibit any transport phenomena too, provided that the localization point is exactly coincident with a local minimum of the potential and the amplitude is big enough.\newline\indent
In both cases we can define $\varphi_{\delta}$ as
\begin{equation}
\varphi_{\delta}(\psi\vert\tau_0) = \int_0^L\psi_{y,\tau_0}K\left(y,x,\tau_0,\tau_0+\delta\vert c_{\tau}=1\right)dy
\end{equation}
where $K(y,x,\tau_0,\tau_0+\delta)$ is the propagator associated with Eq. (\ref{schr}). Figure \ref{figbox} (right panel) gives a graphical explanation of $\phi_{\delta}$ when using $v_1^p$.

\subsubsection{Definition of $\varphi_{\pm}$}
\label{pmpar}
The definition of $\varphi_{\pm}(\cdot\vert\tau)$ is based on the fact that a localized matter wave can experience a periodic potential as the single well where the mass is concentrated, provided that the potential amplitude is big enough and that $x_{loc}$ is near enough to the local minimum $x_{min}$ of the potential. In case of a symmetric well, the symmetry of the eigenstates is well defined and there is a time interval $\epsilon/2$ after which it holds that
\begin{equation}
\psi\left(x,\tau+\frac{\epsilon}{2}\right)\simeq\psi\left(x+2(x_{min}-x_{loc}),\tau\right)
\label{semiosc}
\end{equation}
Let us suppose that 
\begin{equation}
x_{loc}(\tau)=x_{min}-\frac{\Delta}{4}-\delta_x \quad \textrm{with }\delta_x\ll\Delta
\end{equation}. 
From Eq. (\ref{semiosc}) we have
\begin{equation}
x_{loc}\left(\tau+\frac{\epsilon}{2}\right)=2x_{min}-x_{loc}(\tau)=x_{min}+\frac{\Delta}{4}+\delta_x
\end{equation}
 Applying an instantaneous $\pi$ phase shift to the periodic potential leads to a displacement of the local minimum $x_{min}\rightarrow x'_{min} = x_{min}+\Delta/2$. Now we have
\begin{equation}
x_{loc}\left(\tau+\frac{\epsilon}{2}\right)=x'_{min}-\frac{\Delta}{4}+\delta_x
\end{equation}
and after one more $\epsilon/2$ time lapse we obtain
\begin{equation}
x_{loc}(\tau+\epsilon)=2x'_{min}-x_{loc}\left(\tau+\frac{\epsilon}{2}\right)=x_{loc}(\tau)+\Delta
\end{equation}
So we made the localized matter wave to travel a distance $\Delta$ by applying two periodic potential in antiphase and with big amplitude. In Section \ref{robustsec} we investigate the conditions under which this displacement can be iterated preventing $DPR(\tau)$ from rising beyond an acceptable level. The discussion above leads to a definition of $\varphi_{\pm}$:
\begin{eqnarray}
\varphi_{\pm}(\psi\vert\tau_0) &=& \int_0^L\psi_{y,\tau_0}K\left(y,x,\tau_0,\tau_0+\delta\vert c_{\tau}=c^{\pm}_{\tau}\right)dy\\
c^{+}_{\tau}&=&2\cdot\mathbf{1}_{\{\tau\in[\tau_0,\tau_0+\frac{\epsilon}{2}]\}}+3\cdot\mathbf{1}_{\{\tau\in[\tau_0+\frac{\epsilon}{2},\tau_0+\epsilon]\}}\nonumber\\
c^{-}_{\tau}&=&3\cdot\mathbf{1}_{\{\tau\in[\tau_0,\tau_0+\frac{\epsilon}{2}]\}}+2\cdot\mathbf{1}_{\{\tau\in[\tau_0+\frac{\epsilon}{2},\tau_0+\epsilon]\}}\nonumber
\end{eqnarray}
\begin{figure}[hbt]
\includegraphics[width=1\columnwidth]{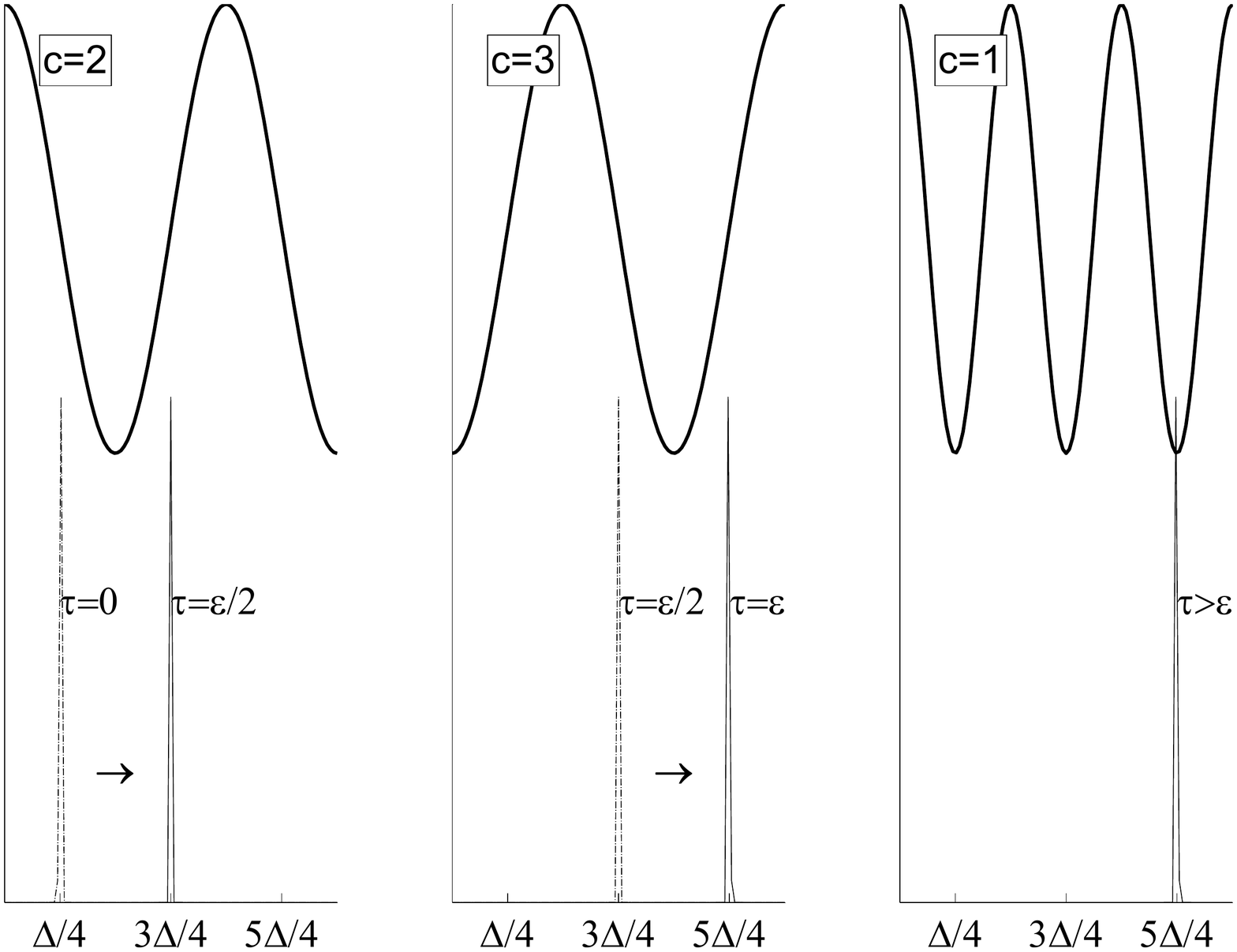} 
\caption{Graphical explanation of the effects described in sections \ref{mempar} and \ref{pmpar}.}
\label{figbox}
\end{figure}
\FloatBarrier
Figure \ref{figbox} (left and central panels) gives a graphical explanation of $\phi_{+}$.
\subsubsection{$\varphi_{\pm}$ near to the boundaries}
\label{bound}
The periodic potential in the torus is translation invariant with respect to the transformation $x\mapsto x\pm\Delta$. This implies that the mechanism described in paragraph \ref{pmpar} holds in any portion of the system in the same way. This leads to eq. (\ref{pmtorus}). Eq. (\ref{eqntorus}) originates from the fact that $x_{loc}$ can be moved only by $\Delta$ long steps and so there are only $N$ allowed positions where $x_{loc}$ can be found, as shown in Figure \ref{figbox} (right panel).
\newline\indent
On the other hand, the box boundary condition has no translation invariance and the localized matter wave is reflected by the infinite potential walls. As shown in Figure \ref{alternative} $x_{loc}$ is shifted by $\Delta/2$ near to the boundaries, because $x_{loc}(\tau/2)=x_{loc}(\tau)$. This fact implies that there are $2N$ allowed positions where $x_{loc}$ can be found. We can enumerate these positions in the order that we obtain by iterative application of $\phi_+(\cdot\vert\tau)$ operator to $\psi_{x,\tau_0}$. The resulting order is described by eq. (\ref{eqnbox}). An example is shown in Figure \ref{figbox} (left panel).
\begin{figure}[hbt]
\includegraphics[width=1\columnwidth]{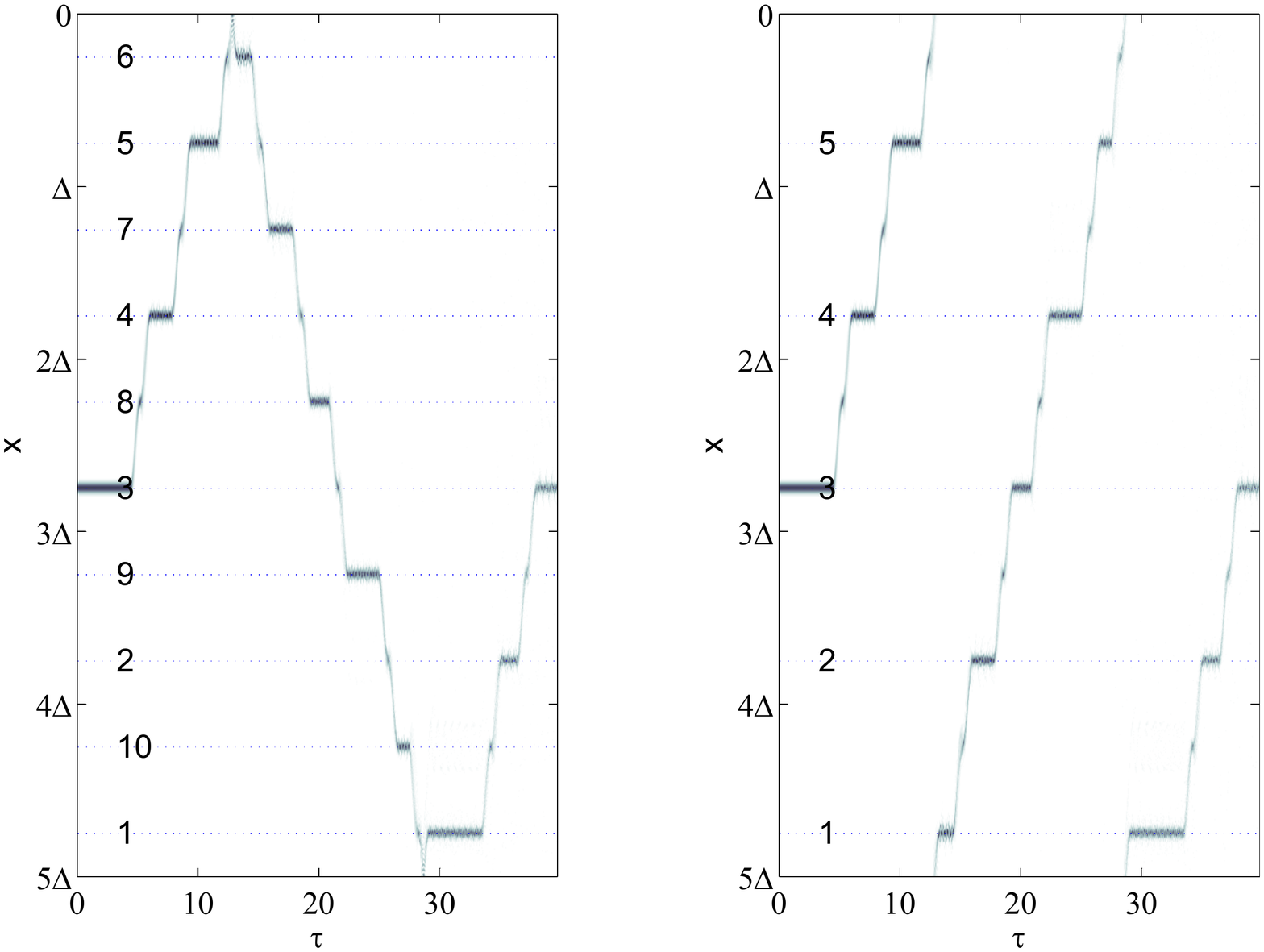} 
\caption{Density plot of $\vert\psi(x,\tau)\vert^2$, using the specifications shown in Figure \ref{figexample_v} (periodic case $v_1^p$). We applied $\varphi_\delta\circ\varphi_+$ ten times, considering a random different $\delta$ value per application. The same pattern is simulated considering both box (left panel) and boundary (right panel) conditions.}
\label{figbox}
\end{figure}
\section{Multi-state bit under various settings}
\label{robustsec}
\begin{figure}[hbt]
\includegraphics[width=1\columnwidth]{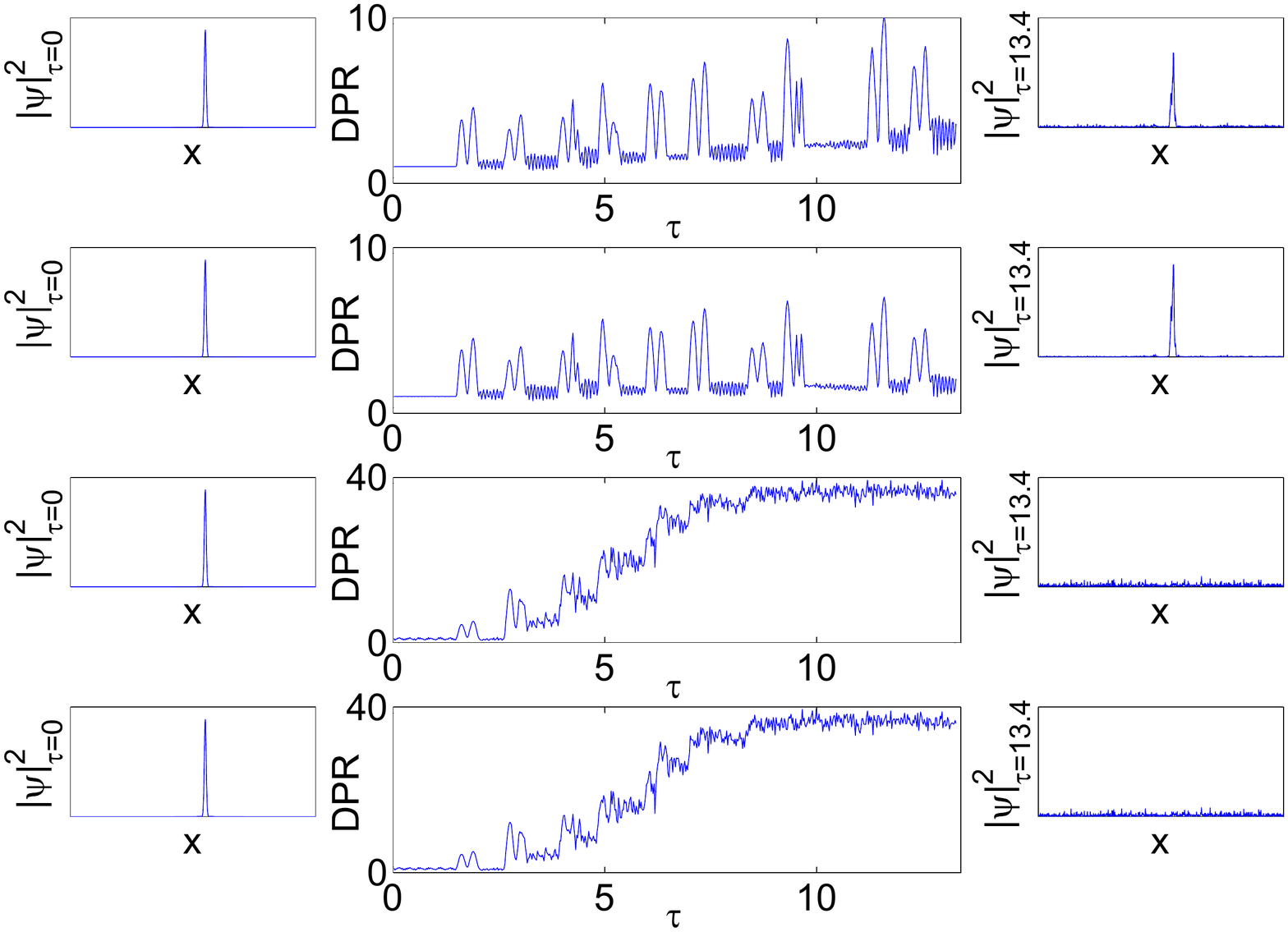} 
\caption{Time evolution of four different versions of the multi-bit system under the same pattern used in Figure \ref{figbox}. The left panels show the initial density $\vert\psi(x,\tau=0)\vert^2$. The central panels plot $DPR(\tau)$ during the evolution of the system. The right panels show the final state of the system $\vert\psi(x,\tau=13.4)\vert^2$. From the top to the bottom: periodic case ($v_1^p$), $a=\beta=0$; periodic case ($v_1^p$) , $\alpha=0.01 \beta=-15$;  disordered case ($v_1^s$) , $\alpha=\beta=0$; disordered case ($v_1^s$), $\alpha=0.01$ and $\beta=-15$.  All the considered versions have in common the following settings: $N=5$; $a_i=120$ $(i=0,\dots,3)$; box boundary conditions.}
\label{figdpr}
\end{figure}

\begin{figure}[hbt]
\includegraphics[width=1\columnwidth]{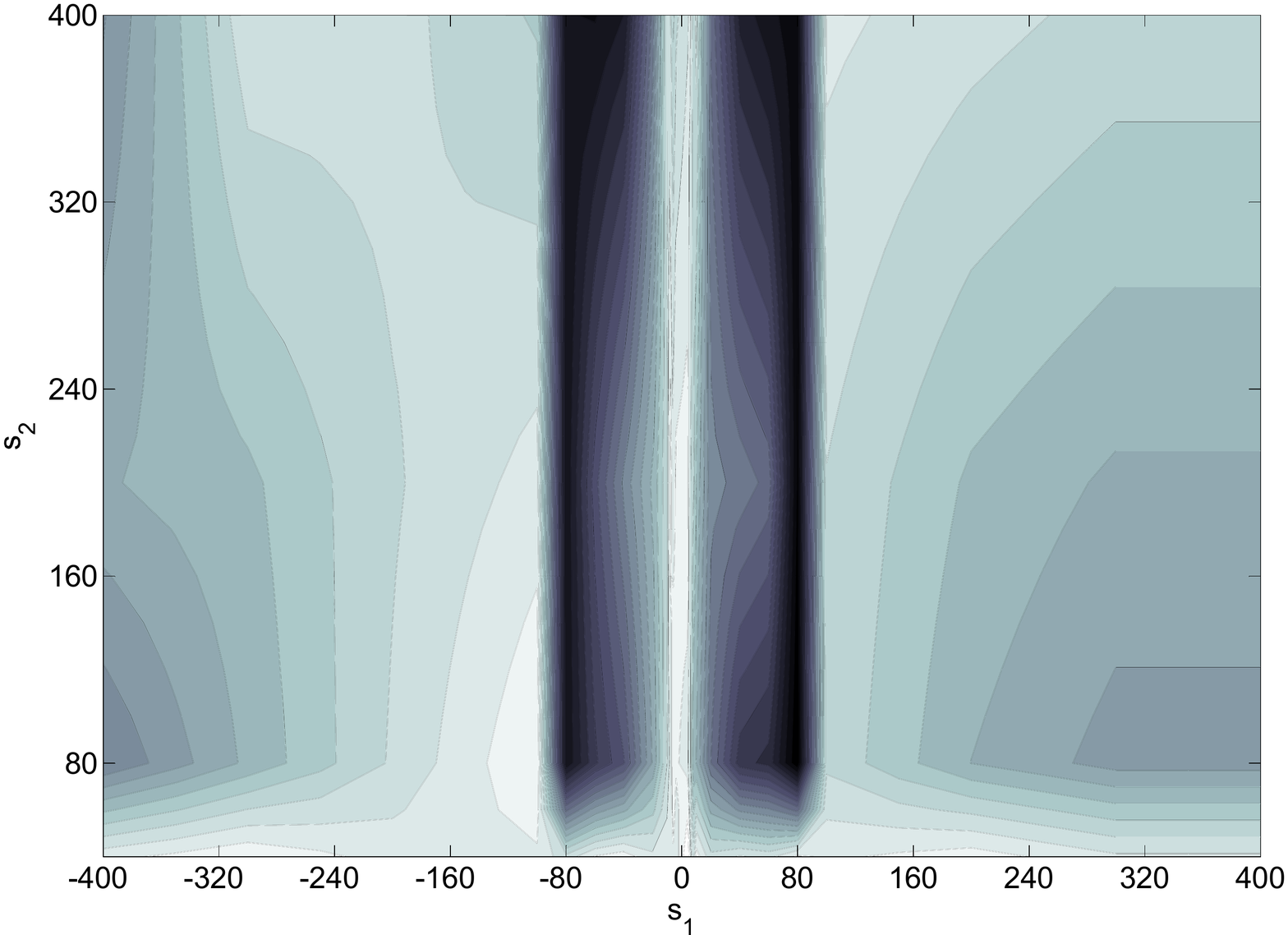} 
\caption{$\textrm{DPR}\lbrack\psi\rbrack$ after five applications of $\varphi_\delta\circ\varphi_{+}$ ($N=7$; disordered potential $v_1^d$; $\alpha=\beta=0$). Darker areas corresponds to lower DPR values.}
\label{figdpr2}
\end{figure}

\begin{figure}[hbt]
\includegraphics[width=1\columnwidth]{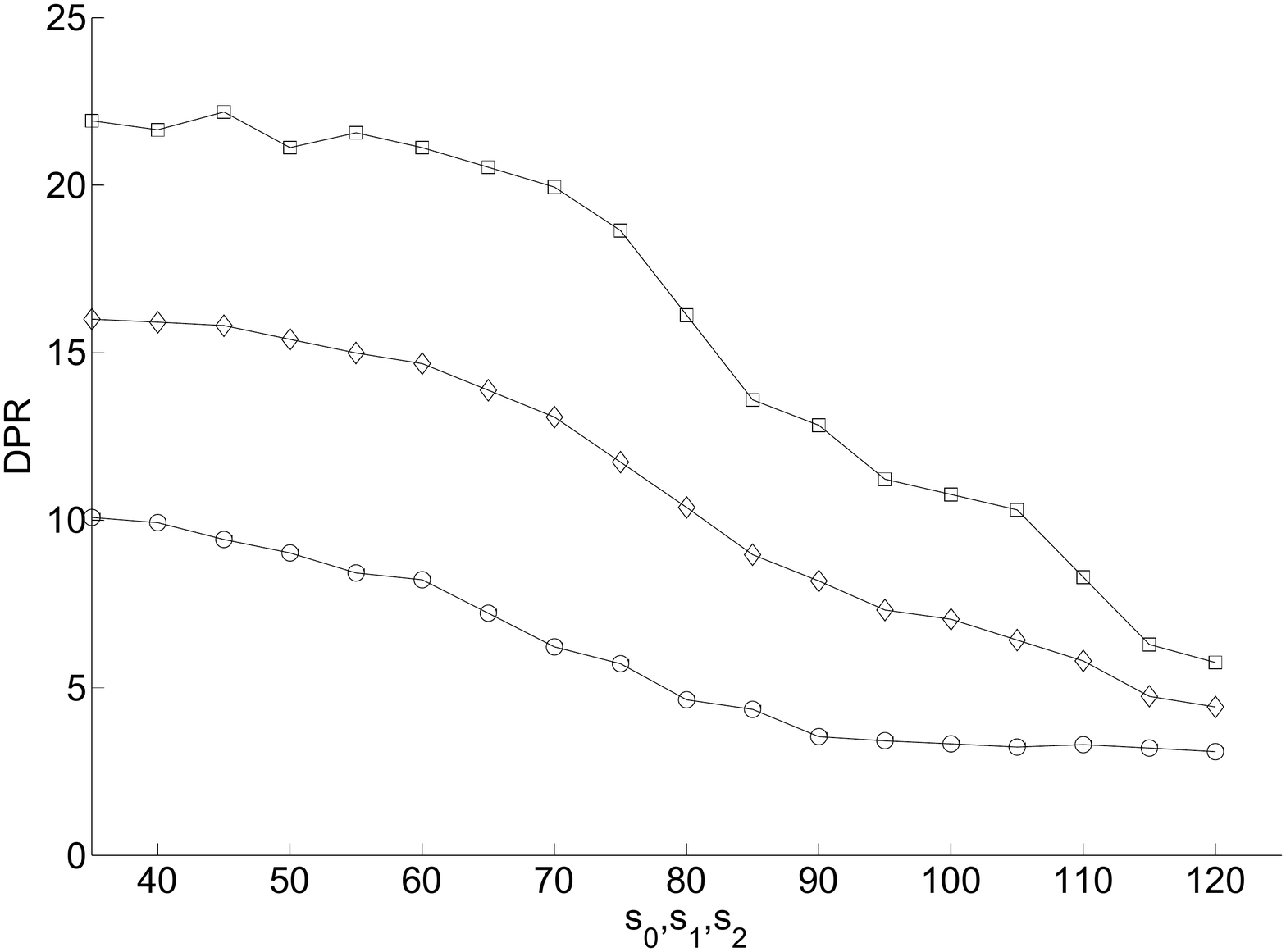} 
\caption{Final $DPR$ after the evolution of the system ($\alpha=\beta=0$; box boundary conditions; periodic case $v_1^p$) using the same pattern described in Figure \ref{figbox} and applied also in Figure \ref{figdpr}. We tested $N$ ranging from $5$ (circles) to $7$ (squares) and $s_0=s_1=s_2$ ranging from $35$ to $120$.}
\label{alternative}
\end{figure}

In this section we verify the possibility of the system to be used as a multi-state bit under different parameters choices. We compare the results using DPR, defined in section \ref{dprpar}. Furthermore, we are interested in finding a stable multi-bit example which can be also feasible in laboratory.\newline\indent
We have compared the two considered potential choices $v_1^d$ and $v_1^p$, using them to storage information in the system. In case we choose to maintain $c_{\tau}=1$ constant, both of them are equally good in keeping the matter wave localized over time. $v_1^p$ results to be better than $v_1^d$ when $\phi_{\pm}$ is  applied repeatedly to the system. This is the case when we want to write information to be stored in the system. An example can be seen in figure \ref{figdpr} (first row and third row panels).
When using $v_1^d$, $\vert s_1\vert$ must be big enough to keep $\psi(x,\tau)$ localized over time, but not big enough to cause fragmentation phenomena. If the matter wave is fragmented, $DPR$ increases when $\psi(x,\tau)$ is forced to oscillate ($c_{t}>1$). Figure \ref{figdpr2} shows an example of optimal $\vert s_1\vert$ level of  when using $v_1^d$. On the other hand, when using $v_1^p$, $s_1$ can be chosen arbitrarily high without fragmenting the matter wave. This is the reason why $v_1^p$ leads to a more stable multi-bit behavior than $v_1^d$.\newline\indent When considering self-interaction, we observed an increased stability (lower DPR over time) choosing $\beta<0$, especially if using $v_1^p$. An example of this fact is shown in figure \ref{figdpr} (second row and fourth row panels). Intuitively, $\beta>0$ decreases the stability of the system. \newline\indent
Moreover, all the potential amplitudes must be higher at increasing $N$ values, in order to keep $\psi(x,\tau)$ confined in a local fluctuation of the potential when the total number of fluctuations $N$ is bigger. An example of this fact is shown in figure \ref{alternative}.
Considering these results, a feasible experimental setting that allows to observe a stable ten states multi-bit could be the following: $^{39}K$ elongated BEC ($10^4$ atoms) under box boundary conditions; system dimensions $300\mu m \times 30\mu m$; periodic $v_1^p$ potential; $s_1=s_2=s_3\geq 90$; $N=5$. Please see appendix \ref{appnpse} (and references therein) for further details.

\section{Summary}
\label{secend}
We have developed a technique to change and preserve the position of a localized matter wave. This behavior is directly applicable to obtain a multi$-$state memory device. This system can be obtained using optical potentials already available in laboratory. The multi-bit behavior can be observed under multiple parameters choices and we have suggested a fully specified multi-bit which could be realized nowadays. Given that BECs and optical potentials are currently investigated from a quantum information perspective, this work opens the possibility of turning the same BEC from a q-bit into a classical multi-state bit and vice versa in the future.  

\section{Acknowledgements} 
Fruitful discussions with M. Modugno are acknowledged.
\newpage
\appendix
\section{1D NPSE in our units}
\label{appnpse}
We consider the 1D NPSE equation \cite{Salasnich2002}, which describe the dynamics of an elongated BEC:
\begin{eqnarray}
i\hbar\frac{\partial}{\partial t}\psi &=& \left[-\frac{\hbar^2}{2 m}\frac{\partial^2}{\partial x^2}+V+\frac{gN_a}{2\pi a_{\perp}}\frac{\vert\psi\vert^2}{\sqrt{1+2a_sN_a\vert\psi\vert^2}}\right. \\
&+&\left.\frac{\hbar\omega_{\perp}}{2}\left(\frac{1}{\sqrt{1+2a_sN_a\vert\psi\vert^2}}+\sqrt{1+2a_sN_a\vert\psi\vert^2}\right)\right]\psi,\nonumber
\label{npse}
\end{eqnarray}
with $a_{\perp}=\sqrt{\frac{\hbar}{m\omega_{\perp}}}$ and $g=\frac{4\pi\hbar^2a_s}{m}$.
Let us introduce the following quantities
\begin{eqnarray}
\tau&:=&\frac{E_{\xi} t}{\hbar}\\
\label{eqtau}
\alpha&:=&\frac{\hbar\omega_{\perp}}{2}\frac{1}{E_{\xi}}=\frac{\xi^2}{a_{\perp}^2}\\
\label{eqalpha}
\beta&:=&2a_sN_a
\label{eqbeta}
\end{eqnarray}
Moreover, it holds that
\begin{equation}
\frac{gN_a}{2\pi a_{\perp}}\frac{1}{E_{\xi}} =\frac{4\pi\hbar^2a_s}{m}\frac{N_a}{2\pi a_{\perp}}\frac{2m\xi^2}{\hbar^2}=2\alpha\beta
\label{eqg}
\end{equation}
multiplying Eq. (\ref{npse}) by $\frac{1}{E_{\xi}}$ and replacing eq. (\ref{eqtau}, \ref{eqalpha}, \ref{eqbeta}, \ref{eqg}) and choosing $\xi=1$ as the spatial unit, we have
\begin{equation}
i\frac{\partial}{\partial \tau}\psi=\left[-\frac{\partial^2}{\partial x^2}+v+\alpha\left(\frac{2\beta\vert\psi\vert^2+1}{\sqrt{1+\beta\vert\psi\vert^2}}+\sqrt{1+\beta\vert\psi\vert^2}\right)\right]\psi.
\label{npse2}
\end{equation}
We simulate a $^{39}K$ condensate with tunable attractive interactions. The following parameters values are accessible to the experiments (see \cite{clement2006} and \cite{Roati2007} amongst others): $\xi\simeq1\mu m$, $N_a\simeq 10^4$, $L\simeq 300\xi$, $a_{\perp}\simeq 30\xi$, $0\geq a_s\gtrsim-7.5\cdot 10^{-4}\xi$. This leads to $\alpha\simeq  10^{-2}$ and $\beta\in[-15, 0]\xi$.


\end{document}